\documentstyle[amssymb,preprint,aps]{revtex}
\begin{document}
\title{Kaluza-Klein gravity in the context of modern embedding theories of the
spacetime}
\author{J. J. Figueiredo and C. Romero}
\address{Departamento de F\'{i}sica, Universidade Federal da Para\'{i}ba, C.Postal\\
5008, 58051-970 Jo\~{a}o Pessoa, Pb, Brazil\\
E-mail: cromero@fisica.ufpb.br}
\maketitle

\begin{abstract}
We review and discuss the original Kaluza-Klein theory in the framework of
modern embedding theories of the spacetime, such as the recent induced
matter approach. We show that in spite of their seeming similarity they
constitute rather distinct proposals as far as their geometrical structure
is concerned.
\end{abstract}

\section{Introduction}

Perhaps one of the most recurrent themes of contemporary theoretical physics
is the idea of unification of the fundamental forces of nature. From the
pioneering paper by Nordstr$\emptyset $m \cite{Nordström}, in 1914 (before
the completion of general relativity), through the old and modern versions
of Kaluza-Klein theory \cite{kaluza,kaluza1,kaluza2}, supergravity \cite
{deser}, superstrings \cite{superstrings}, to the more recent braneworld
scenario \cite{rs1,rs2}, induced-matter \cite{overduin,book} and M-theory 
\cite{duff}, physicists have explored the odd, however apparently fruitful
thought, that unification might finally be achieved if one accepts that
space-time has more than four dimensions.

Among all these higher-dimensional theories, one of them, the induced-matter
theory (also referred to as space-time-matter theory \cite{overduin,book})
stands out for its closeness to the Einstein's project of considering matter
and radiation as manifestations of pure geometry\cite{Einstein}. Indeed, the
gist of the whole theory is to assert that, by embedding the ordinary
space-time into a five-dimensional vacuum space, it is possible to describe
the macroscopic properties of matter in geometrical terms. Picking up
several examples of cosmological and gravitational models, the theory shows
how to interpret the energy-momentum tensor corresponding to some standard
matter configurations in terms of the geometry of the five-dimensional
vacuum space. The mathematical consistency of this theory (as it was later
realised \cite{Romero}) is entirely built on a beautiful and powerful
theorem of differential geometry known as the Campbell-Magaard theorem \cite
{Campbell}.

The induced-matter theory is sometimes given the name of Kaluza-Klein
non-compactified gravity, since Klein's compactness condition is dropped
from the basic assumptions of the theory. Nonetheless, it appears that this
identification with Kaluza-Klein theory is not entirely justifiable, for the
two theories have, as we shall see later, distinct geometrical features.

\section{The Kaluza-Klein approach}

\bigskip

In the original Kaluza's approach, first developed in 1919 \cite{kaluza,Pais}%
, it is assumed that the space-time $M^{5}$\ is five-dimensional, but with
all fields being independent of the fifth dimension. Later, in 1926, Klein
extended this hypothesis by admiting that the fields could depend on the
extra coordinate which, however, was assumed to be compact ( with a radius
of the order of the Planck length). In this scheme, the five-dimensional
space-time $M^{5}$ is viewed as possessing the topology $M^{4}$X $S^{1}$,
and the fifth coordinates $y$ is regarded as periodic ( compactness
condition). (Under these assumptions, it is admitted that our normal
perception of space-time is never able to see this extra dimension.)

The way Kaluza-Klein theory geometrizes the electromagnetic field, thereby
unifying it with gravity, is by postulating that Einstein's general theory
of relativity holds in five dimensions and that in these dimensions our
Universe is empty. The next step is to suppose that it is possible to find a
special coordinate system, such that we can make the following (4+1)-split
of the five-dimensional metric 
\begin{equation}
\overline{g}_{ab}=%
%TCIMACRO{
%\binom{g_{\mu \upsilon }+\phi ^{2}A_{\mu }A_{\nu }\ \ \ \ \ \ \ \ \phi ^{2}A_{\mu }}{\ \ \ \phi ^{2}A_{\upsilon }\ \ \ \ \ \ \ \ \ \ \ \ \ \ \ \ \phi ^{2}\ } }%
%BeginExpansion
{g_{\mu \upsilon }+\phi ^{2}A_{\mu }A_{\nu }\ \ \ \ \ \ \ \ \phi ^{2}A_{\mu } \choose \ \ \ \phi ^{2}A_{\upsilon }\ \ \ \ \ \ \ \ \ \ \ \ \ \ \ \ \phi ^{2}\ }%
%EndExpansion
\label{kaluzametric}
\end{equation}
where $A_{\mu }$ may be regarded as a 4-dimensional vector, and $\phi $ is a
scalar field, the role of which is to be determined later \footnote{%
Throughout the paper we shall be using the following convention, notation
and units. Greek indices run over 0,1,2,3, and small Latin indices run over
0,1,2,3,4,. We take the signature of the five-dimensional space to be (+ - -
- -) . We also shall make the identifications $x=x^{\mu },x^{4}=y,$ $%
x^{a}=(x,y)$, and work in units such that $c=1$.}. Now, due to the
compactness condition,\ all the fields may be expanded in Fourier series:

\[
g_{\mu \upsilon }(x,y)=\sum_{n=-\infty }^{n=\infty }g_{\mu \upsilon
}^{\left( n\right) }(x)e^{iny/l},\text{ }A_{\mu }(x,y)=\sum_{n=-\infty
}^{n=\infty }A_{\mu }^{\left( n\right) }(x)e^{iny/l},\text{ }\phi
(x,y)=\sum_{n=-\infty }^{n=\infty }\phi ^{\left( n\right) }(x)e^{iny/l} 
\]
where $l$ denotes the ``length'' of the fifth dimension and the subscript $n$
refers to the $n$th Fourier mode. Since $l$ is supposed to be very small ($%
\sim 10^{-35}$m) , in low energy regime only the the mode $n=0$ is usually
retained in the above expansions. This amounts to effectively regarding $%
g_{\mu \upsilon }$, $A_{\mu }$ and $\phi $ as independent of the fifth
coordinate $y$.

The Einstein vacuum field equations in five-dimensions are 
\begin{equation}
G_{ab}=0  \label{einstein}
\end{equation}
or, equivalently, 
\begin{equation}
R_{ab}=0  \label{ricci}
\end{equation}
where $G_{ab}=R_{ab}-R\overline{g}_{ab}/2$ denotes the Einstein tensor in
five dimensions, $R_{ab}$ and $R$ is the Ricci tensor and the curvature
scalar, respectively, all these quantities being calculated with the
five-dimensional metric tensor $\overline{g}_{ab}$ $\footnote{%
We are adopting the following definition for the Ricci tensor: $%
R_{ab}=\partial _{c}\Gamma _{ab}^{c}-\partial _{b}\Gamma _{ac}^{c}+\Gamma
_{ab}^{c}\Gamma _{cd}^{d}-\Gamma _{ad}^{c}\Gamma _{bc}^{d}$.}$.

We now come to the interesting thing about Kaluza-Klein theory. By a
straightforward calculation one can show that it is possible to separate the
equation (\ref{einstein}) into the following set: 
\begin{equation}
^{(4)}G_{\mu \upsilon }=\frac{\phi ^{2}}{2}T_{\mu \nu }-\frac{1}{\phi }\left[
\nabla _{\mu }(\partial _{\upsilon }\phi )-g_{\mu \upsilon }\square \phi %
\right]  \label{einstein1}
\end{equation}
\begin{equation}
\nabla ^{\mu }F_{\mu \upsilon }=-3\frac{\partial ^{\mu }\phi }{\phi }F_{\mu
\upsilon }  \label{4-potential}
\end{equation}
\begin{equation}
\square \phi =\frac{\phi ^{3}}{4}F_{\mu \upsilon }F^{\mu \upsilon }
\label{scalar}
\end{equation}
where the quantities $^{(4)}G_{\mu \upsilon }=^{(4)}R_{\mu \upsilon
}-^{(4)}Rg_{\mu \upsilon }/2$, $^{(4)}R_{\mu \upsilon }$ and $^{(4)}R$ are
calculated with the ``four-dimensional'' metric $^{(4)}g_{\mu \upsilon
}\equiv g_{\mu \upsilon }(x)$ and, thereby, interpreted as the
four-dimensional ``Einstein tensor''. On the other hand, $T_{\mu \nu }=$ $%
g_{\mu \upsilon }F^{a\beta }F_{\alpha \beta }/4-F_{\mu }^{\alpha }F_{\nu
\alpha }$ plays the role of the electromagnetic energy-momentum tensor, with 
$F_{\mu \upsilon }\equiv \partial _{\mu }A_{\nu }-\partial _{\nu }A_{\mu }$.

At this point two comments are in order. Firstly, let us just note that when 
$A_{\mu }=0$ the above equations are formally equivalent to the vacuum
Brans-Dicke field equations \cite{brans-dicke} with $w=0$ ( the equation (%
\ref{4-potential})\ being reduced to an identity). Secondly, we may be
tempted to say that if we take $\phi =1$, then we are led to the
Einstein-Maxwell equations for a radiating field. However, this is no quite
so, for although the equations (\ref{einstein1}) and (\ref{4-potential})
support such interpretation, the equation (\ref{scalar}) for the scalar
field introduces the very undesirable constraint $F_{\mu \upsilon }F^{\mu
\upsilon }=0$. There is, nevertheless, a clever mathematical trick to
overcome this difficult. Instead of trying to deduce the Einstein-Maxwell
theory directly from (\ref{einstein}), let us take advantage of the
lagrangian formalism. We start by writing the action describing a
source-free space-time (pure gravity) in five dimensions. The action
describing this system is given by 
\begin{equation}
^{(5)}S=-\frac{1}{16\pi \overline{G}}\int d^{5}x\sqrt{-\overline{g}}^{(5)}R
\label{action}
\end{equation}
where $\overline{G}$ is a ``five-dimensional gravitational constant''.\ It
can be shown that if we split up the five-dimensional curvature scalar $%
^{(5)}R$\ in terms of $\ ^{(4)}R$ , the fields $A_{\mu }$, $\phi $ and their
derivatives, we end up, after integrating (\ref{action}) with respect to the
compact coordinate $y$, with the ``four-dimensional'' action 
\begin{equation}
^{(4)}S=-\int d^{4}x\sqrt{-g}\phi \left( \frac{^{(4)}R}{16\pi G}+\frac{1}{4}%
\phi ^{2}F_{\mu \upsilon }F^{\mu \upsilon }+\frac{2}{3\kappa ^{2}}\frac{%
\partial _{\mu }\phi \partial ^{\mu }\phi }{\phi ^{2}}\right)
\label{4-action}
\end{equation}
where we have substituted $\kappa ^{2}=16\pi G$, and $G=\overline{G}/l$ is
identified to the Newtonian gravitational constant in four dimensions. Now,
if one takes $\phi =const$, then one recovers the action corresponding to
the gravitational field interacting with a radiating electromagnetic field.
In this sense, we can say that the electromagnetic field appears exclusively
from the geometry of the the five-dimensional space-time. Generalizations of
this procedure in order to \bigskip incorportate other kinds of fields, such
as the strong and weak nuclear fields, constitutes what are now called
modern Kaluza-Klein theories \cite{kaluza1,kaluza2}.

Let us assume that the invariance group $G_{5}$\ of the admissible
coordinate transformations is the direct product $G_{4}$X$G_{1}$, where $%
G_{4}$ is the manifold general group of $M^{4}$ ( i.e. the set of general
transformations $x^{\prime \mu }=x^{\prime \mu }(x)$ )\ and $G_{1}$ is
defined as the set of coordinate transformations of the type $y^{\prime
}=y+f(x)$. A straightforward calculation shows then that under $S_{1}$ the
functions $A_{\mu }$ transform as$\ A_{\mu }^{\prime }=$ $A_{\mu }-\frac{%
\partial f}{\partial x^{\mu }}$, which means that $G_{1}$ is a geometric
version of the gauge group of electromagnetism. On the other hand, it is
easy to verify that under $G_{4}$ the functions $\phi ,$ $A_{\mu }$\ and $%
g_{\mu \nu }$ behave as scalar, vector, and tensor fields, respectively.

\bigskip Before concluding this brief review let us raise the question of
whether Kaluza-Klein theory should be regarded as an embedding theory, as it
is often suggested in the literature. Now, an embedding theory \ assumes, as
a first principle, that our ordinary space-time $M^{4}$ corresponds to some
hypersurface embedded in a five-dimensional manifold $M^{5}$. An example of
an embedding theory is the recently proposed brane-world theory \cite
{rs1,rs2}, where our observable Universe is viewed as a four-dimensional
hypersurface embedded in a five-dimensional anti-de Sitter manifold (the
so-called bulk). Another example of embedding theory is the induced-matter
theory (although in this case, other formulations, such as the foliation
approach, are also possible \cite{fifthforce} ). It turns out that regarding
Kaluza-Klein theory as an embedding theory is rather problematic, if not
impossible. We shall return to this point later.

\section{\protect\bigskip The induced-matter approach}

The induced-matter theory made its first appearance in the early 1990s. \ It
postulates that our ordinary space-time $M^{4}$ may be viewed as a
four-dimensional hypersurface embedded in a five-dimensional Ricci-flat
space $M^{5}$. In this proposal, put forward by Wesson and colaborators (
for a detailed description and references, see \ \cite{book}\ ), vacuum
(4+1)-dimensional Einstein equations give rise to (3+1)-dimensional
equations with sources. By choosing appropriate embeddings, it is possible,
for instance, to derive the standard Friedmann-Robertson-Walker cosmological
models \cite{kalligas}, de Sitter vacuum space-times \cite{poncedeleon},
etc. As we have seen, the original version of Kaluza-Klein theory assumes as
a postulate that the fifth dimension is compact. In the case of the
induced-matter theory this condition has been completely dropped . A second
basic tenet of the theory is that all classical macroscopic physical
quantities, such as matter density and pressure, should be given a
geometrical interpretation. In this way, it is proposed that the classical
energy-momentum tensor, which enters the right-hand side of the usual
four-dimensional Einstein equations can, in principle, be generated by pure
geometrical means. In other words, it is claimed that geometrical curvature
``induces'' matter in four dimensions ( this is why it is called
``induced-matter theory''), and to an observer in the ordinary
four-dimensional space-time the extra dimensions appear as the matter source
for gravity.

Thus the induced-matter approach assumes that the fundamental
five-dimensional space $M^{5}$, in which our usual spacetime $M^{4}$ is
embedded, is a solution of the five-dimensional vacuum Einstein equations (%
\ref{ricci}). The geometry of $M^{5}$ is given by $ds^{2}=\overline{g}%
_{ab}dx^{a}dx^{b}$, where the metric tensor $\overline{g}_{ab}=\overline{g}%
_{ab}(x,y)$ is written as 
\begin{equation}
\overline{g}_{ab}=%
%TCIMACRO{
%\binom{\overline{g}_{\mu \upsilon }\ \ \ \ \ \ \ \ \ \ \ \ \ \ \ \ 0}{\ \ \ 0\ \ \ \ \ \ \ \ \ \ \ \ \ \ \ \ \epsilon \phi ^{2}\ } }%
%BeginExpansion
{\overline{g}_{\mu \upsilon }\ \ \ \ \ \ \ \ \ \ \ \ \ \ \ \ 0 \choose \ \ \ 0\ \ \ \ \ \ \ \ \ \ \ \ \ \ \ \ \epsilon \phi ^{2}\ }%
%EndExpansion
\label{imtmetric}
\end{equation}
where $\varepsilon =\pm 1$, and now the Kaluza-Klein electromagnetic
potential has been removed, as all non-diagonal components $\overline{g}%
_{4\alpha }$ of the metric have been set to zero. It is important to note
here that we cannot obtain (\ref{imtmetric}) from (\ref{kaluzametric}) by a
coordinate transformation allowed by the Kaluza-Klein's invariance group $%
G^{5}$, unless in the trivial case when the electromagnetic potentials are
of the type $A_{\mu }=\partial _{\mu }\Lambda $ ( pure gauge ), with $%
\Lambda (x)$ any differentiable function. Thus, the geometrization of the
electromagnetic field in the induced-matter approach, unlike Kaluza-Klein
thoery, is not carried out by putting $A_{\mu }$ directly in the
five-dimensional metric. Indeed, in the induced-matter approach the
electromagnetic field will manifest itself through its energy-momentum
tensor, which must be generated by the fifth dimension in conjunction with
the dependence of the scalar and metric fields on the fifth coordinate $y$.
In this procedure, the ``Maxwell equations'' (\ref{4-potential})\ will be
repaced by another set of four equations involving only the metric, the
scalar field and their derivatives. Let us go into the details.

The five-dimensional Ricci tensor in terms of the five-dimensional
Christoffel symbols is given by 
\begin{equation}
R_{ab}=\partial _{c}\Gamma _{ab}^{c}-\partial _{b}\Gamma _{ac}^{c}+\Gamma
_{ab}^{c}\Gamma _{cd}^{d}-\Gamma _{ad}^{c}\Gamma _{bc}^{d}  \label{fivericci}
\end{equation}
If we identify $\overline{g}_{\mu \upsilon }$ with the metric of the our
ordinary four-dimensional space time $M^{4}$, then by putting $a\rightarrow
\alpha ,b\rightarrow \beta $ in (\ref{fivericci}) we get an equation which
relates the four-dimensional components of the $R_{ab}$ to \ $%
^{(4)}R_{\alpha \beta }$, is the four-dimensional Ricci tensor calculated
with $\overline{g}_{\alpha \beta }$. It is not difficult to verify that \cite
{de Leon} 
\begin{equation}
R_{\alpha \beta }=^{(4)}R_{\alpha \beta }+\Gamma _{\alpha \beta
,4}^{4}-\Gamma _{\alpha 4,\beta }^{4}+\Gamma _{\alpha \beta }^{\lambda
}\Gamma _{\lambda 4}^{4}+\Gamma _{\alpha \beta }^{4}\Gamma _{4d}^{d}-\Gamma
_{\alpha \lambda }^{4}\Gamma _{\beta 4}^{\lambda }-\Gamma _{\alpha
4}^{d}\Gamma _{\beta d}^{4}
\end{equation}
where $^{(4)}R_{\alpha \beta }$ is the four-dimensional Ricci tensor
calculated with $\overline{g}_{\alpha \beta }$, which is to be identified to
the observed four-dimensional metric itself. We then can show that the
five-dimensional vacuum Einstein equations (\ref{ricci})\ can be written,
separetely, in the following way: 
\begin{equation}
^{(4)}G_{\alpha \beta }=\kappa T_{\alpha \beta }  \label{einstein2}
\end{equation}
where $\kappa $ is the Einstein constant and $\ T_{\alpha \beta }$ is
interpreted as the energy-momentum tensor of the ordinary four-dimensional
matter, and is given explicitly by 
\begin{eqnarray}
\kappa ^{(4)}T_{\alpha \beta } &=&\frac{\phi _{\alpha ;\beta }}{\phi }-\frac{%
\varepsilon }{2\phi ^{2}}\left[ \frac{\phi _{,4}g_{\alpha \beta ,4}}{\phi }%
-g_{\alpha \beta ,44}+g^{\lambda \mu }g_{\alpha \lambda ,4}g_{\beta \mu ,4}-%
\frac{g^{\mu \nu }g_{\mu \nu ,4}g_{\alpha \beta ,4}}{2}\right]
\label{energy-momentum tensor} \\
&+&\frac{\varepsilon }{2\phi ^{2}}\frac{g_{\alpha \beta }}{4}\left\{
g_{,4}^{\mu \nu }g_{\mu \nu ,4}+\left( g^{\mu \nu }g_{\mu \nu ,4}\right)
^{2}\right\}  \label{energy-momentum}
\end{eqnarray}

\bigskip

\begin{equation}
\varepsilon \phi \square \phi =-\frac{g_{,4}^{\lambda \beta }g_{\lambda
\beta ,4}}{4}-\frac{g^{\lambda \beta }g_{\lambda \beta ,44}}{2}+\frac{\phi
,_{4}g^{\lambda \beta }g_{\lambda \beta ,4}}{2\phi }  \label{scalar1}
\end{equation}
which may be viewed as an equation for a scalar field $\phi ;$ and

\begin{equation}
P_{\alpha ;\beta }^{\beta }=0  \label{conservation}
\end{equation}
an equation that has the appearance of a conservation law, where $P_{a\beta
} $ is defined by 
\begin{equation}
P_{\alpha \beta }=\frac{1}{2\sqrt{g_{44}}}\left( g_{\alpha \beta
,4}-g_{\alpha \beta }g^{\mu \nu }g_{\mu \nu ,4}\right)
\end{equation}
Therefore, we see that the five-dimensional vacuum field equations can be
splitted into three parts: equations (\ref{einstein2}),(\ref{scalar1}) and (%
\ref{conservation}).

There are at least two distinct versions of the induced-matter theory, which
will be referred to as the{\it \ foliation }and {\it \ embedding approaches}%
, and which lead to different results as far as the dynamics of particles
and fields is concerned (\cite{fifthforce}). They are defined as follows:

i) The foliation approach makes use of a congruence of the vector field $V=%
\frac{\partial }{\partial y}$, defined in $M^{5}$, and implicitly assumes
that the equations governing the four-dimensional observed physical laws are
in a way ''projections'' of five-dimensional equations onto the foliation of
hypersurfaces $\left\{ \Sigma \right\} $ ( defined by $y=const$\ )
orthogonal to $V.$ In this approach the geometry of the four-dimensional
space-time is determined by inducing the metric of $M^{5}$ on the leaves, so
that $^{(4)}g_{\mu \upsilon }=g_{\mu \nu }(x,y)$ (note in this case the
dependence of the metric tensor on the extra coordinate $y$ ).

ii) In the embedding approach it is also assumed that $M^{5}$ can be
foliated by a set of \ hypersurfaces $\left\{ \Sigma \right\} $ orthogonal
to a vector field $V.$ However, here the geometry of the ordinary space-time 
$M^{4}$\ is not supposed to be determined by the entire foliation, but by a
particular leaf $\Sigma ^{4}$ (say, $y=0$ ), selected from the set $\left\{
\Sigma \right\} $, on which a metric tensor is induced by the embedding
manifold $M^{5}.$ In \ this approach the geometry is determined in terms of
quantities which are defined exclusively in $\Sigma ^{4}$ and, in
particular, the metric of $M^{4}$ is given by $^{(4)}g_{\mu \upsilon
}=g_{\mu \nu }(x,y=0)$.

\section{Is Kaluza-Klein gravity an embedding theory?}

Finally, we are able to answer the question addressed in section II: Is
Kaluza-Klein theory to be regarded as an embedding theory? In other words,
considering $\overline{g}_{ab}$ as given by equation (\ref{kaluzametric}),
is it possible to find a hypersurface $\Sigma _{4}$ of $M^{5}$, such that
the metric induced on $\Sigma _{4}$ by $\overline{g}_{ab}$ is $g_{\mu \nu
}(x)$ ? Let us try to answer this question by supposing that such a
hypersurface exists and may be parametrized by an equation of the type $%
y=f(x)$.

The line element of $M^{5}$ may be written as 
\begin{equation}
dS^{2}=\overline{g}_{ab}dx^{a}dx^{b}=(g_{\mu \upsilon }+\phi ^{2}A_{\mu
}A_{\nu })dx^{\mu }dx^{\nu }+2\phi ^{2}A_{\mu }dx^{\mu }dy+\phi ^{2}dy^{2}
\label{dskaluza}
\end{equation}
\bigskip Now, substituting $dy=\partial _{\mu }fdx^{\mu }$ into (\ref
{dskaluza}) we get 
\[
ds^{2}=(g_{\mu \upsilon }+\phi ^{2}A_{\mu }A_{\nu }+2\phi ^{2}A_{\mu
}\partial _{\nu }f+\phi ^{2}\partial _{\mu }f\partial _{\nu }f)dx^{\mu
}dx^{\nu } 
\]
In order for the induced metric to be $g_{\mu \nu }(x)$ we must have $A_{\mu
}=-\partial _{\mu }f$ . But then we see that $A_{\mu }$ must be a pure
gauge, so in this case we have no electromagnetic field in (\ref{dskaluza}).
Therefore, we conclude that, in the framework of Kaluza-Klein, our
four-dimensional space-time $M^{4}$ may be viewed as a hypersurface $\Sigma
_{4}$ embedded in the five dimensional vacuum space $M^{5}$ only if the
electromagnetic field is ``switched off''. This mathematical fact may
explain why the electromagnetic four-potential $A_{\mu }$ must be dropped
from the very beginning in the induced-matter formalism, if the latter is to
be regarded as an embedding theory.

\section{Acknowledgement}

The authors would like to thank CNPq for financial support.

\end{document}